\documentclass[osajnl,twocolumn,showpacs,superscriptaddress,10pt]{revtex4-1} 
\usepackage{amsmath,amssymb,graphicx}
\usepackage{color}
\begin{document}

\title{Frequency correlations in reflection from random media}

\author{Angelika Knothe}
\author{Thomas Wellens}\email{Corresponding author: thomas.wellens@physik.uni-freiburg.de}
\affiliation{Physikalisches Institut, Albert-Ludwigs-Universit\"at Freiburg, Hermann-Herder-Str. 3, D-79104 Freiburg, Germany}

\begin{abstract}
We present a theoretical study of frequency correlations of light backscattered from a random scattering medium. This statistical quantity provides insight into the dynamics of multiple scattering processes accessible both, in theoretical and experimental investigations. For frequency correlations between field amplitudes, we derive a simple expression in terms of the path length distribution of the underlying backscattering processes. In a second step, we apply this relation to describe frequency correlations between intensities in the regime of weak disorder. Since, with increasing disorder strength, an unexplained breakdown of the angular structure of the frequency correlation function has recently been reported in experimental studies, we explore extensions of our model to the regime of stronger disorder. In particular, we show that closed scattering trajectories 
tend to suppress the angular dependence of the frequency correlation function.
\end{abstract}


\maketitle 

\section{Introduction}

Correlations in frequency space have proven to be an important tool for analyzing the dynamics of multiple scattering processes of waves in disordered media. 
For fixed scattering particles, i.e., in a setup where the frequency dependence is not influenced by eventual movements of the scatterers~\cite{maret_multiple_1987}, the second order frequency correlation function of multiply scattered field amplitudes 
can be represented as the Fourier transform of the time-of-flight distribution \cite{genack_relationship_1990}. 
This relation has been exploited, e.g., in early experimental investigations 
\cite{genack_optical_1987, garcia_crossover_1989, genack_relationship_1990,van_albada_observation_1990},
studying intensity-intensity correlations of light or microwaves transmitted through random samples in dependence of the system parameters.
In the same spirit, more recent experimental studies of intensity-intensity correlations in transmission demonstrated how these can be used to extract dynamical transport parameters like delay times~\cite{genack_statistics_1999,  van_tiggelen_delay-time_1999} or the diffusion constant \cite{muskens_method_2009} in transmission experiments through disordered media. 

This picture of the immediate 
connection between frequency correlations and the path lengths of the scattering sequences becomes even richer when considering the case of reflection from a random medium where, due to different contributing scattering processes as compared to the transmission scenario, the path length distribution depends crucially on the backscattering angle $\vartheta$. This fact manifests itself in the effect of coherent backscattering \cite{vinogradov73,kuga_retroreflectance_1984, wolf_weak_1985,banakh87,banakh13}, leading to a
sharp angular cone of the backscattered intensity around the direction $\vartheta=0$  corresponding to exact backscattering. Consequently, this dependency of the path length distribution on the backscattering angle leads to a characteristic angular peak of the width of the frequency correlation function \cite{muskens_angle_2011}.
In a previous work on
reflection from two- and three-dimensional random wave guides~\cite{schomerus_localization-induced_2001},
the narrow angular peak  that results in the diffusive regime of weak disorder could not be resolved, due to the discrete 
scattering channel geometry. On the other hand, however, this work explicitly reports
on effects of the transition between the localized and the diffusive regime on the 
delay time statistics.
In particular, a significantly broader distribution of delay times is observed in the localized regime (i.e., for strong disorder) than in the diffusive regime (i.e., for weak disorder).
In a related discussion concerning the physics of the localization transition, correlation properties of radiation reflected from random media, both, in angular~\cite{aubry_recurrent_2014} or frequency~\cite{hildebrand_observation_2014} space, recently have been observed to undergo drastic changes as the disorder becomes sufficiently strong. Being directly related to the dynamical features of the underlying scattering processes, 
frequency correlations are considered promising for obtaining a better understanding of the change of scattering mechanisms and propagation properties at (or close to) the Anderson localization transition. 
In that context, 
recent experimental results~\cite{muskens_angle_2011} 
indicate that the characteristic angular peak of the width of the frequency correlation function mentioned above displays a drastic breakdown in the regime of stronger disorder near the localization threshold.
The fact that this effect has so far not been explained theoretically provides the main motivation for the present paper.

In this work, we focus on the case of reflection from a random scattering medium and present detailed theoretical studies of the intensity-intensity  correlations in the frequency domain. This represents in two ways a completion and an extension of the previous works described above: On the one hand, we enrich the knowledge about frequency correlations in reflection as compared to the case of transmission by tracing back the properties observed in experiment to the nature of the contributing scattering processes for a backscattering geometry. 
In particular, we show that, for the case of a laser beam with large transverse width $\rho\gg\ell$ (where $\ell$ denotes the mean free path) and for backscattering angles $\vartheta>0$, frequency correlations are also affected by propagation {\em outside} the scattering medium. We derive an analytic expression for this dependency, which allows to map the frequency correlation function between field amplitudes onto the path length distribution of multiple scattering trajectories {\em within} the scattering medium (and vice versa). Furthermore,
by investigating the impact of varying the amount of disorder in the system, we discuss how frequency correlations might help to shed more light on possible changes occurring at the localization transition on the level of the underlying scattering mechanisms.

After building up the theoretical frame in Sec.~\ref{sec:TheoFrame}, we show in Sec.~\ref{sec:MF_IntensityProp} how to describe 
frequency correlations between multiply scattered field {\em amplitudes} in the regime of weak disorder, thereby developing a simple relation between the frequency correlation function and the distribution of path lengths of the multi-scattering sequences contributing to reflection. 
In Sec.~\ref{sec:FCFweakdisorder}, we apply this relation to treat frequency correlations between field {\em intensities} 
in the regime of weak disorder, providing a theoretical interpretation for previous experimental results \cite{muskens_angle_2011} in this regime. Finally, in Sec.~\ref{sec:Corrections}, in order to investigate the behaviour of frequency correlations near the localization transition, we explore several corrections to the weak disorder approximations applied in Sec.~\ref{sec:FCFweakdisorder}. Whereas corrections to the Gaussian $C^1$-approximation are shown to be negligible in the case of our backscattering setup, we demonstrate that  \textit{recurrent scattering trajectories} \cite{aubry_recurrent_2014} do indeed tend to suppress the angular peak of the frequency correlation function, as it has been observed in experiment \cite{muskens_angle_2011}.

\section{Theoretical Frame}
\label{sec:TheoFrame}

In order to describe the frequency correlations of light reflected from a disordered medium in terms of the underlying multiple random scattering processes, we work in the frame of the following model:
The disorder is described by a random potential $V(\mathbf{r})=\delta n(\mathbf{r})/\langle n \rangle$ describing relative fluctuations of the dielectric constant $n$ with respect to its mean value $\langle n \rangle$ within the region of the scattering medium. The potential $V(\mathbf{r})$ is modeled as a Gaussian random process, which is characterized by its mean value $\langle V(\mathbf{r})\rangle\equiv 0$ and correlation function $\langle V(\mathbf{r})V(\mathbf{r}^{\prime})\rangle$. Here, and in the following, angular brackets denote disorder average over different realization of the disorder. In the regime where the wave length of the scattered wave $\lambda$ is much larger than the correlation length of the disorder potential, the correlation function can be approximated by:
\begin{equation}
\langle V(\mathbf{r})V(\mathbf{r}^{\prime})\rangle=u^{2}\delta(\mathbf{r}- \mathbf{r}^{\prime}),
\label{eq:u2}
\end{equation}
where we introduced the pre-factor $u^2$ characterizing the scattering strength of the disorder. In scalar approximation, a wave $^{\omega}\psi(\mathbf{r})$ with wave number 
$^{\omega}k=\frac{\omega}{c}$
propagating in the presence of the disorder potential satisfies the scalar Helmholtz equation 
\begin{equation}
\Big[  \nabla^2+\,^{\omega} k^2 \big(1+ V(\mathbf{r})    \big)    \Big]\;^{\omega}\psi  (\mathbf{r}) = \rho(\mathbf{r}) ,\label{eq:helmholtz}
\end{equation}
where $\rho(\mathbf{r})$ denotes a given source distribution. 

Instead of treating a special solution for one particular realization of the disorder, we will be interested in average properties. First,
we describe the average propagation between the points $\mathbf{r}$ and $\mathbf{r}^{\prime}$ in space, i.e., we consider the average Green function $^{\omega}G(\mathbf{r}-\mathbf{r}^\prime): =  \langle  ^{\omega}G_V(\mathbf{r},\mathbf{r}^\prime) \rangle$ obtained by ensemble averaging the Green function $^{\omega}G_V(\mathbf{r},\mathbf{r}^\prime)$ associated to Eq.~(\ref{eq:helmholtz}) for one special configuration $V$ of the disorder. The average propagator can be written as \cite{akkermans_mesoscopic_2011}  
\begin{equation}
^{\omega}G({\mathbf{r},\mathbf{r}^{\prime}})=-\frac{1}{4\pi}\frac{e^{\,i \;^{\omega}k|{\mathbf{r}- \mathbf{r}^{\prime}}|}}{|{\mathbf{r}- \mathbf{r}^{\prime}}|}\; e^{-\frac{|{\mathbf{r}- \mathbf{r}^{\prime}}|}{2\ell}},
\label{eqn:G}
\end{equation}
which exhibits, compared to the vacuum case $V\equiv0$, where the propagator is given by a spherical wave, additionally a term describing exponential damping. The damping constant $\ell$ is referred to as the \textit{scattering mean free path}, which describes the average distance between two successive scattering events. With $\ell$ as the characteristic length scale of the disorder, we further introduce the \textit{disorder parameter} $k\ell$ classifying the amount of disorder in the system by comparing the wavelength of the scattered wave to the scattering mean free path. Hence, the disorder is said to be \textit{weak} for large values $k\ell\gg1$ of the disorder parameter, whereas the strength of the disorder increases for decreasing values for $k\ell$. For weak disorder, the mean free path turns out as $\ell=4\pi/(k^4u^2)$~\cite{akkermans_mesoscopic_2011}. Moreover, in this regime, scattering of the average wave {\em intensity} can be described by ladder and crossed diagrams (see Sec.~\ref{sec:MF_IntensityProp}) which, in turn, are constructed from average Green functions, Eq.~(\ref{eqn:G}), and scattering events defined by the potential correlation function, Eq.~(\ref{eq:u2}), introduced above.

\section{Frequency correlations between field amplitudes}
\label{sec:MF_IntensityProp}

\begin{figure}
\includegraphics[width=0.25\textwidth]{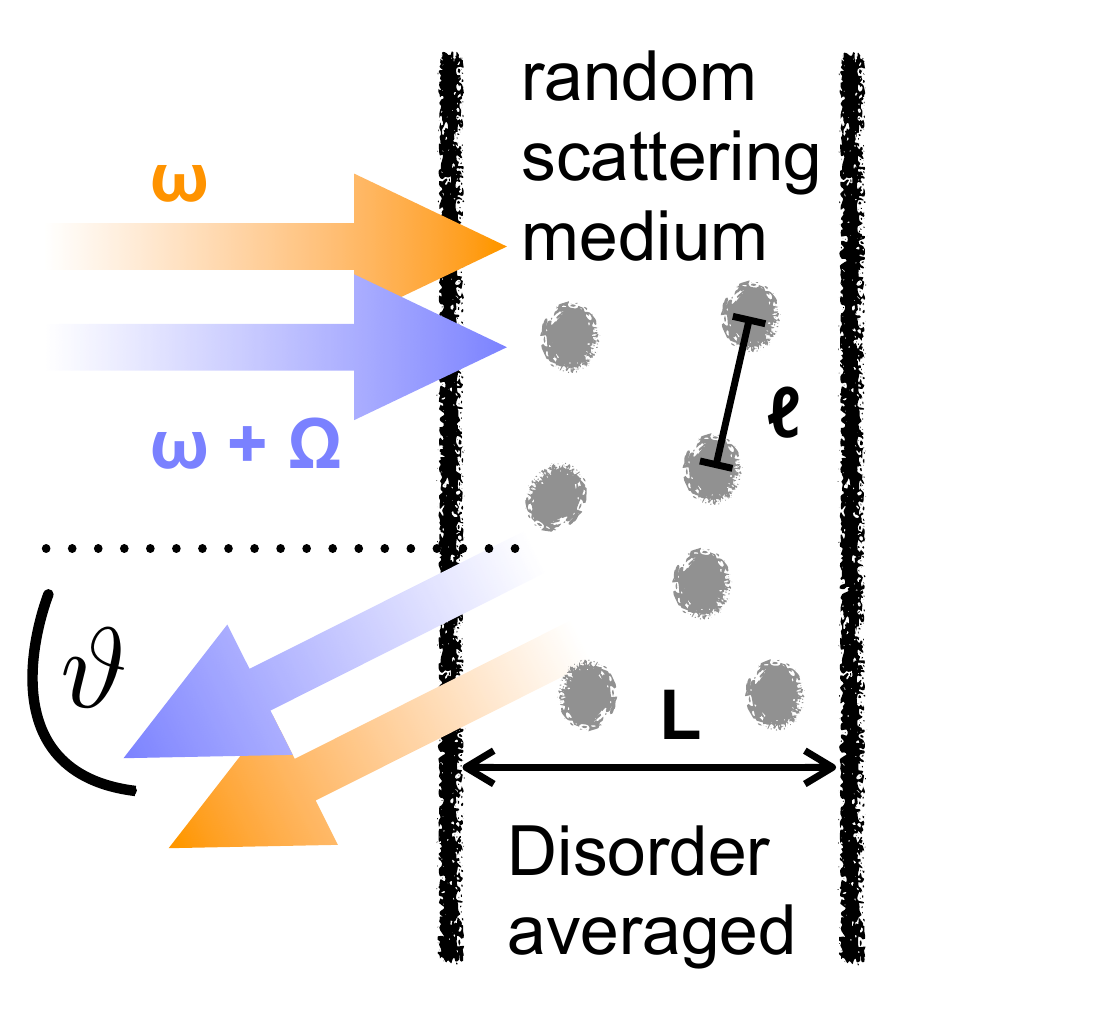}
\caption{Scattering setup under consideration: Light of two different frequencies $\omega$ (orange) and $\omega+\Omega$ (blue) is backscattered under the backscattering angle $\vartheta$ from a random scattering medium where the disorder is characterized by the scattering mean free path $\ell$ as the average distance between two successive scattering events of the multi-scattering sequence. Disorder average is taken over several different realizations of the random potential. \label{fig:Setup}}
\end{figure}

We aim at describing the statistical properties of light of two different frequencies $\omega$ and $\omega+\Omega$ which is backscattered from a random scattering medium. For the sample, we choose a slab geometry of thickness $L$, within which the disorder is characterized by the scattering mean free path $\ell$. The setup under consideration is sketched in Fig.~\ref{fig:Setup}. In the following, we consider purely elastic scattering and we neglect the effects of absorption or internal reflections on the boundaries.

\begin{figure}
\vskip5pt\includegraphics[width=0.38\textwidth]{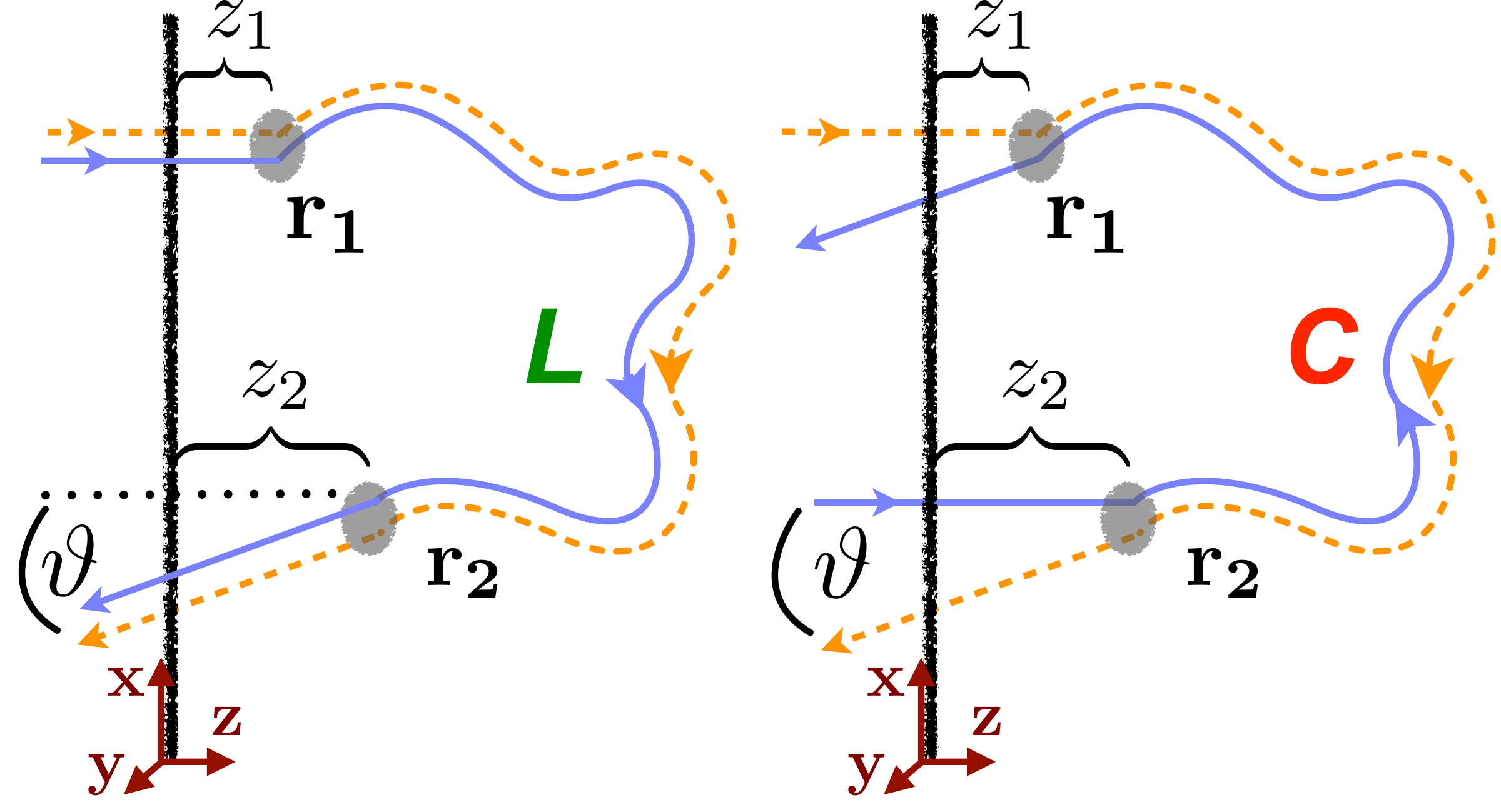}
\caption{Scattering processes yielding minimal dephasing between the wave $^{\omega+\Omega}\psi_{out}$ at frequency $\omega+\Omega$ (solid lines) and its complex conjugate $^{\omega}\psi_{out}^*$ at frequency $\omega$ (dashed lines) as they are taken into account to compute the field correlation function $F_\vartheta(\Omega)$, see Eq.~(\ref{eq:I}), within the approximation of weak disorder: Ladder (L) and crossed (C) process, where the same scatterers are visited in the same or in the reversed order, respectively. 
\label{fig:LundC}}
\end{figure}

The first quantity we focus on is the \textit{field correlator} 
\begin{equation}
F_\vartheta(\Omega)=\langle ^{\omega+\Omega}\psi_{out}(\vartheta)\;^{\omega}\psi^{*}_{out}(\vartheta) \rangle,\label{eq:I}
\end{equation}
which can be seen as a generalized intensity composed of the two outgoing fields $^{\omega+\Omega}\psi_{out}(\vartheta)$ and $^{\omega}\psi_{out}(\vartheta)$ which are scattered into the same direction parametrized by the scattering angle $\vartheta$, but differ in frequency by $\Omega$. 
To compute 
this quantity, we apply the following approximation, as it is commonly used in the treatment of disorder averaged intensity propagation in the case of weak disorder: We assume  only those interference processes which lead to a minimum dephasing between the wave and its complex conjugate to yield non-negligible contributions on average.
Minimal phase shift is achieved either if the wave and the complex conjugate visit the same scatterers in the same order (ladder propagation, L) or in reversed order (crossed propagation, C). The equivalence of ladder and crossed processes in media exhibiting time-reversal symmetry for $\Omega=0$ leads to the famous \textit{coherent backscattering effect} \cite{vinogradov73,kuga_retroreflectance_1984,wolf_weak_1985,banakh87,banakh13}. 
Within this \textit{approximation of minimal dephasing}, the 
ladder and crossed
scattering processes shown in Fig.~\ref{fig:LundC} have to be taken into account in order to compute $F_\vartheta(\Omega)$:
\begin{equation}
F_\vartheta(\Omega)=F^{(L)}_\vartheta(\Omega)+F^{(C)}_\vartheta(\Omega),\label{eq:I_tot}
\end{equation}
where $F^{(L)}_\vartheta(\Omega)$ and $F^{(C)}_\vartheta(\Omega)$
are constructed out of the following components: The incident field $^{\omega}\psi_{in}(\mathbf{r})$ of amplitude $\psi_{0}$ is described by a Gaussian beam of width $\rho$ and wave vector $^{\omega}\mathbf{k}_{in}=\frac{\omega}{c}\mathbf{e}_z$ perpendicular to the $x$-$y$ plane as the surface of incidence:
\begin{equation}
^{\omega}\psi_{in}(\mathbf{r})=\psi_{0}\,e^{-\frac{z}{2\ell}}\,e^{-\frac{(\mathbf{r}^{\bot})^2}{2\rho^2}}\,e^{i\frac{\omega}{c}z},
\label{eqn:phi_in}
\end{equation}
where $\mathbf{r}^{\bot}=\sqrt{x^2+y^2}$ denotes the transversal component of the vector $\mathbf{r}=(x,y,z)$, i.e., its projection onto the plane of incidence.

The outgoing radiation is described in the far field ($R\gg r$, where $\mathbf{R}$ refers to the position of the detector and $\mathbf{r}$ to a position within the scattering medium) using Eq.~(\ref{eqn:G}) in Fraunhofer approximation  \cite{akkermans_mesoscopic_2011}:
\begin{eqnarray}
^{\omega}G_{out}(\mathbf{r}, \mathbf{R})&= & \frac{1}{4\pi}\frac{e^{i\,^{\omega}k|\mathbf{R}-\mathbf{r}|}}{|\mathbf{R}-\mathbf{r}|}\, e^{-\frac{z}{2\ell\cos\vartheta}}\nonumber\\
& \approx &  e^{-i\frac{\omega}{c}(x\sin\vartheta-z\cos\vartheta)}  \frac{e^{i\frac{\omega}{c}R}}{4\pi R}  e^{-\frac{z}{2\ell\cos\vartheta}},
\label{eqn:G_out}
\end{eqnarray}
where we used $\mathbf{R}=R \mathbf{e}_{out}$ and $^{\omega}\mathbf{k}_{out}=\,^{\omega}k\,\mathbf{e}_{out}=\frac{\omega}{c}(\cos\vartheta,0, -\sin\vartheta)$ to parametrize the outgoing scattering direction. Hence, $z/\cos\vartheta$ gives the distance travelled by the field inside the medium from the point of the last scattering event to the boundary surface, cf. Fig.~\ref{fig:LundC}.
The wavy lines in Fig.~\ref{fig:LundC} connecting the points $\mathbf{r}_1$ and $\mathbf{r}_2$ indicate the ladder (L) or crossed (C) propagation process of the average intensity between these two points, which we describe with the \textit{mixed frequency intensity propagators} $^{\Omega}P_{L}(\mathbf{r}_1,\mathbf{r}_2)$ and $^{\Omega}P_{C}(\mathbf{r}_1,\mathbf{r}_2)$, respectively, where the superindex $\Omega$ indicates a frequency shift $\Omega$ between the two propagating fields. Using the following notion of a \textit{single-step propagator} 
\begin{align}
\nonumber^{\Omega}P_{0}(\mathbf{r}_1,\mathbf{r}_2)&=\, ^{\omega+\Omega}G(\mathbf{r}_1,\mathbf{r}_2)\; ^{\omega}G^{*}(\mathbf{r}_1,\mathbf{r}_2)\\
&=\frac{1}{(4\pi)^2|\mathbf{r}_1-\mathbf{r}_2|^2}e^{-\frac{|\mathbf{r}_1-\mathbf{r}_2|}{\ell}}e^{i\frac{\Omega}{c}|\mathbf{r}_1-\mathbf{r}_2|},
\label{eqn:P0}
\end{align}
to describe a single scattering step between points $\mathbf{r}_1$ and $\mathbf{r}_2$ in space in terms of the average Green function given in Eq.~(\ref{eqn:G}), the intensity propagator can be written in the following, self-consistent integral representation:
\begin{equation}
^\Omega P_L(\mathbf{r}_1,\mathbf{r}_2)=u^2\delta(\mathbf{r}_1-\mathbf{r}_2) + \,  u^2\int d\mathbf{r}\, ^\Omega P_{0}(\mathbf{r}_1,\mathbf{r}) ^\Omega P_L(\mathbf{r},\mathbf{r}_2),
\label{eqn:Prop}
\end{equation}
where the term $u^2\delta(\mathbf{r}_1-\mathbf{r}_2)$ represents single scattering.
Remembering the symmetry of the Green function under exchange of the spatial arguments, $ ^{\omega}G(\mathbf{r}_1,\mathbf{r}_2)=\, ^{\omega}G(\mathbf{r}_2,\mathbf{r}_1)$, a single crossed scattering step, where one of the scattering paths is reversed compared to its ladder counterpart, is described by the same single-step propagator 
$\nonumber^{\Omega}P_{0}(\mathbf{r}_1,\mathbf{r}_2)$
as a ladder step. Hence, the crossed sequence yields an equivalent propagation process -- except for the fact that single scattering is excluded from the crossed propagator:
\begin{equation}
^\Omega P_C(\mathbf{r}_1,\mathbf{r}_2)= u^2\int d\mathbf{r}\, ^\Omega P_{0}(\mathbf{r}_1,\mathbf{r}) ^\Omega P_L(\mathbf{r},\mathbf{r}_2).
\label{eqn:PropC}
\end{equation}
Using the ingredients of Eqs.~(\ref{eqn:phi_in},\ref{eqn:G_out}) and (\ref{eqn:Prop}), we identify the diagrams of Fig. \ref{fig:LundC} to be described by the following formulae:
\begin{eqnarray}
F^{(L)}_\vartheta(\Omega) & = &  \iint d \mathbf{r}_{1}d \mathbf{r}_{2}\; ^{\omega+\Omega}\psi_{in} (\mathbf{r}_{1})\; ^{\omega}\psi^{*}_{in} (\mathbf{r}_{1})\; ^{\Omega}P_L(\mathbf{r}_{1},\mathbf{r}_{2})\nonumber\\
& & \times \; ^{\omega+\Omega}G_{out} (\mathbf{r}_{2},\mathbf{R}) ^{\omega}G^{*}_{out} (\mathbf{r}_{2},\mathbf{R}),\\
F^{(C)}_\vartheta(\Omega) &  = &  \iint d \mathbf{r}_{1}d \mathbf{r}_{2}\; ^{\omega+\Omega}\psi_{in} (\mathbf{r}_{2}) \;^{\omega}\psi^{*}_{in} (\mathbf{r}_{1})\; ^{\Omega}P_C(\mathbf{r}_{1},\mathbf{r}_{2})\nonumber\\
& & \times \; ^{\omega+\Omega}G_{out}(\mathbf{r}_{1},\mathbf{R}) ^{\omega}G^{*}_{out} (\mathbf{r}_{2},\mathbf{R}).
\label{gamma_C_Om}
 \end{eqnarray}
Inserting Eqs.~(\ref{eqn:phi_in},\ref{eqn:G_out}) for the incoming and outgoing waves, we can explicitly write down the dependence on the frequency shift $\Omega$ and the scattering angle $\vartheta$ as follows:
\begin{eqnarray}
F^{(L)}_\vartheta(\Omega) & = & \frac{e^{i\frac{\Omega}{c} R}}{(4 \pi R)^2} \iint d \mathbf{r}_{1}d \mathbf{r}_{2}\; e^{-\frac{(\mathbf{r}^{\bot}_1)^2}{\rho^2}}e^{-\frac{1}{\ell}(z_1+\frac{z_2}{\cos\vartheta})}\nonumber\\
& & \times \; ^{\Omega} P_L(\mathbf{r}_{1},\mathbf{r}_{2})\;  e^{i\frac{\Omega}{c}\lambda_L(\mathbf{r}_{1}, \mathbf{r}_{2}, \vartheta) },\label{gammaL_Om_tot}\\
F^{(C)}_\vartheta(\Omega) & = & \frac{e^{i\frac{\Omega}{c} R}}{(4 \pi R)^2}\iint d \mathbf{r}_{1}d \mathbf{r}_{2}\; e^{-\frac{(\mathbf{r}^{\bot}_1)^2+(\mathbf{r}^{\bot}_2)^2}{2\rho^2}}e^{-\frac{z_1+z_2}{2\ell}\left(1+\frac{1}{\cos\vartheta}\right)}\nonumber\\
& & \times \; e^{i \; ^{\omega}k\big((x_1-x_2)\sin\vartheta+(z_1-z_2)(1-\cos\vartheta)\big) }\nonumber\\
& & \times \;^{\Omega}P_C(\mathbf{r}_{1},\mathbf{r}_{2})
 e^{i\frac{\Omega}{c}\lambda_C(\mathbf{r}_{1}, \mathbf{r}_{2}, \vartheta) },
 \label{gammaC_Om_tot}
\end{eqnarray}
where we introduced the enter/exit dephasing lengths (see below) for ladder and crossed diagrams:
\begin{eqnarray}
\lambda_L(\mathbf{r}_{1}, \mathbf{r}_{2}, \vartheta) & = & z_1-x_2\sin\vartheta+z_2\cos\vartheta\label{eq:lambda_L},\\
\lambda_C(\mathbf{r}_{1}, \mathbf{r}_{2}, \vartheta) & = & z_2-x_1\sin\vartheta+z_1\cos\vartheta\label{eq:lambda_C}.
\end{eqnarray}
In Eqs.~(\ref{gammaL_Om_tot}, \ref{gammaC_Om_tot}), we recognize (apart from the constant term $e^{i\Omega R/c}$) two terms depending on the frequency shift $\Omega$ and thus giving rise to two different dephasing mechanisms: 

(i) {\em Propagation within the scattering medium} (i.e., between the first and last scattering event), described by the terms $^\Omega P_{L,C}(\mathbf{r}_1,\mathbf{r}_2)$. As evident from Eq.~(\ref{eqn:P0}), a phase factor $e^{i\frac{\Omega}{c}|\mathbf{r}_i-\mathbf{r}_j|}$ is picked up at each single step of the multiple scattering process. For the total sequence, these single phase factors add to a total phase $e^{i\frac{\Omega}{c}\mathcal{L}(\mathbf{r}_1\rightarrow\mathbf{r}_n)}$ accumulated by the propagator $^{\Omega}P_{L,C}(\mathbf{r}_1,\mathbf{r}_n)$, where $\mathcal{L}(\mathbf{r}_1\rightarrow\mathbf{r}_n)=\sum_{i=1}^{n-1}|\mathbf{r}_i-\mathbf{r}_{i+1}|$ denotes the total path length of the scattering path from the initial scatterer at point $\mathbf{r}_1$ to the final scatterer at $\mathbf{r}_n$ (see also Appendix~\ref{sec:pathlength}).

 (ii) {\em Propagation outside the scattering medium} (i.e., before the first and after the last scattering event), described by the terms $e^{i\frac{\Omega}{c}\lambda_{L,C}(\mathbf{r}_{1}, \mathbf{r}_{2}, \vartheta) }$: These terms are defined by the positions $\mathbf{r}_{1,2}$ (in particular their $x$- and $z$-coordinates $x_{1,2}$ and $z_{1,2}$) of the first and the last scattering event, respectively, according to Eqs.~(\ref{eq:lambda_L},\ref{eq:lambda_C}).

These considerations allow us to rewrite Eqs.~(\ref{gammaL_Om_tot},\ref{gammaC_Om_tot}) in the following form:
\begin{equation}
F^{(L,C)}_\vartheta(\Omega)=\iint d\mathcal{L}\,d{\lambda}\,\mathfrak{A}_\vartheta^{(L,C)}(\mathcal{L},\lambda)\, e^{i\frac{\Omega}{c}(\mathcal{L}+\lambda+R)},
\label{eqn:Fourier}
\end{equation}
where $\mathfrak{A}^{(L,C)}_\vartheta(\mathcal{L},\lambda)$ denotes the joint distribution of the path length $\mathcal L$ and the enter/exit dephasing lengths  $\lambda$ for waves scattered into direction $\vartheta$ in the case $\Omega=0$ (see Appendix~\ref{sec:pathlength}). The latter can be obtained by a numerical Monte-Carlo simulation of a random walk within the scattering medium (see Appendix~\ref{sec:montecarlo}). 

Treating the dephasing lengths $\mathcal{L}$ and $\lambda$ approximately as independent variables, which we may assume since $\lambda_{L,C}$ are dominated by the transversal coordinate $x$, 
whereas the total pathlength $\mathcal{L}$ exhibits translational invariance along the plane of incidence, allows us to factorize  the joint distribution
\begin{equation}
\mathfrak{A}^{(L,C)}_\vartheta(\mathcal{L},\lambda) \simeq {A}^{(L,C)}_\vartheta(\mathcal{L})\, {B}^{(L,C)}_\vartheta(\lambda)
\end{equation}
into separate distributions for $\mathcal L$ and $\lambda$, respectively.

The integral over the dephasing length $\lambda$ of the enter/exit process can now be approximated analytically by the following steps:
The transversal coordinates of the initial scatterer  follow a Gaussian distribution of width $\rho$, whereas the longitudinal penetration depth is damped exponentially with damping factor $\ell$, cf. Eq.~(\ref{eqn:phi_in}). Since we assume $\rho\gg\ell$ (which is in accordance with the experimental parameters in  \cite{muskens_angle_2011} and turns out to be a crucial condition to observe effects of crossed propagation processes as we expose in more detail in Sec.~\ref{sec:Corrections}),  $\lambda_{L}$ and $\lambda_{C}$, see Eqs.~(\ref{eq:lambda_L},\ref{eq:lambda_C}), are hence dominated by the transversal coordinate $x$. 
We thus approximate $\lambda_L\simeq\lambda_C\simeq x\sin\vartheta$, where $x=r\sin\varphi$ is expressed in polar coordinates, with uniformly distributed angle $\varphi$ and Gaussian distributed radial coordinate, i.e., $P_t(r)=(\sqrt{\pi}\rho)^{-1}\exp[-r^2/\rho^2]$. The integral over $\lambda$ in Eq.~(\ref{eqn:Fourier}) thus gives rise to the following term:
\begin{align}
\nonumber h(\Omega)&=\int_0^{\infty}dr \,P_t(r) \int_0^{2\pi}\frac{d\varphi}{2\pi}\, e^{i\frac{\Omega}{c}r\sin\varphi\sin\vartheta} \\
&= e^{-\frac{1}{8}(\frac{\Omega}{c})^2\rho^2\sin^2\vartheta} I_0 \left[ \frac{1}{8}\left(\frac{\Omega}{c}\right)^2 \rho^2 \sin^2\vartheta \right],
\label{eqn:h}
\end{align}
where $I_0(x)$ denotes the zeroth modified (hyperbolic) Bessel function of first kind. Thereby, the total correlation function $I_\vartheta(\Omega)$, see Eq.~(\ref{eq:I_tot}), can be expressed in terms of the path length distributions  ${A}^{(L,C)}_\vartheta(\mathcal{L})$ of ladder and crossed propagation sequences  as follows:
\begin{equation}
F_\vartheta(\Omega) \simeq h(\Omega) \int  d\mathcal{L} \, \Big( {A}^{(L)}_\vartheta( \mathcal{L}) + {A}^{(C)}_\vartheta( \mathcal{L}) \Big) e^{i\frac{\Omega}{c}(\mathcal{L}+R)},
\label{eqn:gamma_final}
\end{equation}
Conversely, a measurement of the field correlation function $F_\vartheta(\Omega)$ can be used to determine the distribution of path lengths $\Lambda$ by inverse Fourier transformation of Eq.~(\ref{eqn:gamma_final}). Note that, in the case of a Gaussian beam with large width $\rho$ and backscattering angles $\vartheta>0$, also dephasing due to propagation outside the scattering medium must be taken into account through the factor $h(\Omega)$ in Eq.~(\ref{eqn:gamma_final}).

Comparable experimental measurements  of the path length distribution for  transmission through a random scattering sample have been performed in \cite{rojas_photon_2011}. As the authors point out, however, a measurement of the intensity-intensity correlation function (see Sec.~\ref{sec:FCFweakdisorder}) -- which is insensitive to the phase of the field correlation function $F_\vartheta(\Omega)$ -- does not contain the full statistical information needed to directly recover the path length distribution
without imposing additional assumptions on the latter. It is possible, though, to obtain the required phase information using third order intensity correlations \cite{blount_recovery_1969,webster_temporal_2002, webster_temporal_2003}. 

\section{Frequency correlations between intensities}\
\label{sec:FCFweakdisorder}

\begin{figure}
\includegraphics[width=0.49\textwidth]{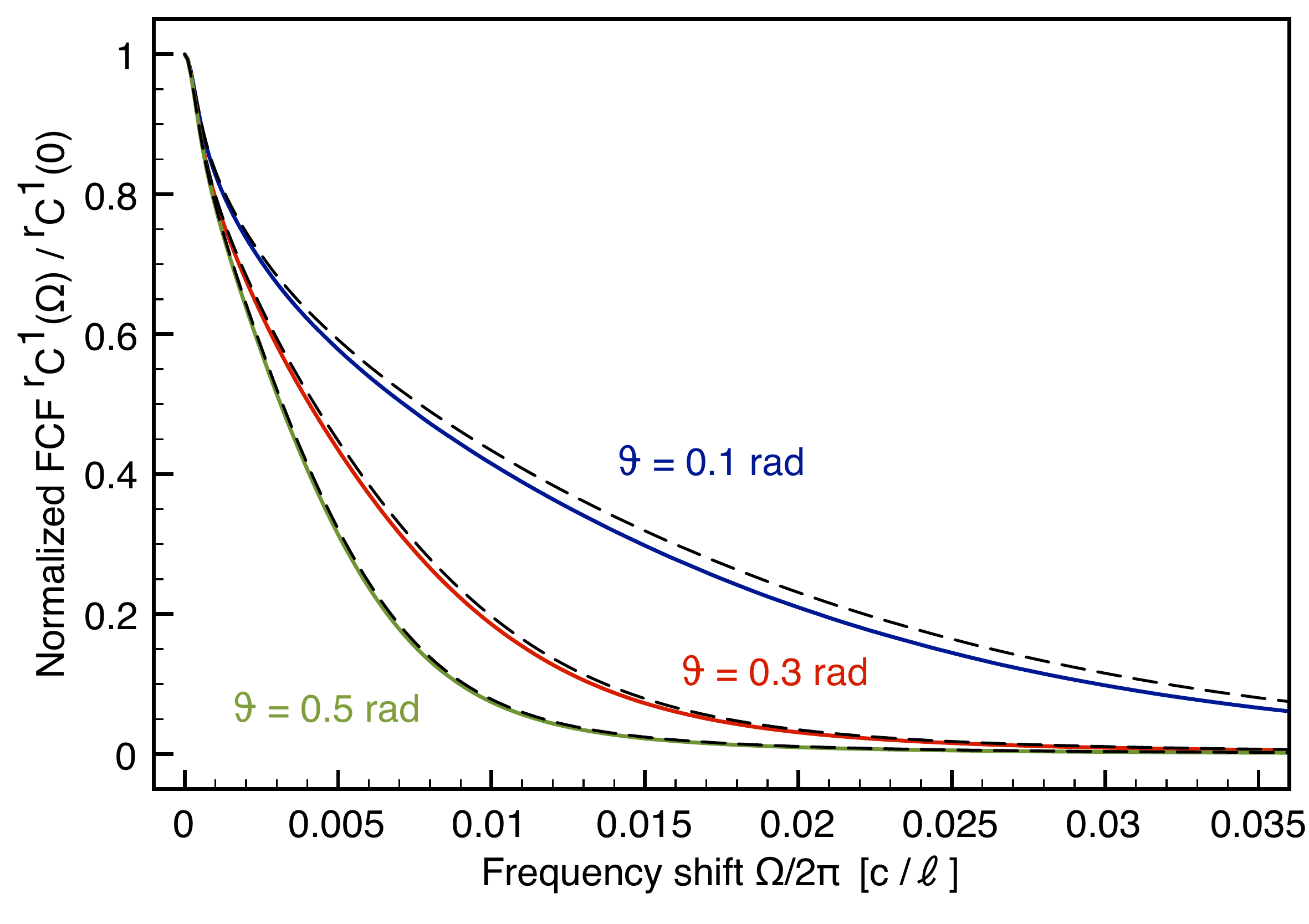}
\caption{We compare the normalized frequency correlation function $C^1_{\vartheta}(\Omega) /C^1_{\vartheta}(0) $ in Gaussian approximation according to Eq.~(\ref{eqn:FCF_gamma}) evaluated in the approximation of minimal dephasing (using ladder and crossed diagrams as shown in Fig.~\ref{fig:LundC}) obtained numerically (solid lines) to the semi-analytic description in terms of the path length distribution, see Eq.~(\ref{eqn:gamma_final}) (dashed lines). Good agreement is obtained for the different backscattering angles $\vartheta=0.1$ (black), $0.3$ (blue) and $0.5$ (red) under study. The system parameters in this plot are given by the disorder parameter $k\ell=5$, the width of the incoming Gaussian beam $\rho=100\ell$ and the optical thickness of the scattering medium $b=L/\ell=30$.\label{fig:Approx}}
\end{figure} 

Since photodetectors are sensitive to the intensity of a light field, most experiments measure frequency correlations between intensities rather than between field amplitudes. In the following, we will consider the normalized quantity:
\begin{equation}
C_{\vartheta}(\Omega) = \frac{\langle ^{\omega+\Omega}I_{out}(\vartheta)\:^{\omega}I_{out}(\vartheta)\rangle}{\langle^{\omega+\Omega}I_{out}(\vartheta)\rangle\;\langle ^{\omega}I_{out}(\vartheta)\rangle}-1,
\label{eqn:FCF}
\end{equation}
where $^\omega I_{out}(\vartheta)=\left|^\omega\psi_{out}(\vartheta)\right|^2$. In order to calculate Eq.~(\ref{eqn:FCF}), we apply the following two approximations valid for
\textit{weakly disordered media}: First, we apply the \textit{Gaussian approximation}, assuming the scattered fields $^{\omega}\psi_{out}(\vartheta)$  to follow Gaussian statistics. As a consequence, the four-field average  in the numerator of Eq.~(\ref{eqn:FCF}) factorizes according to Wick's theorem and can be expressed in terms of two-field contractions:
\begin{equation}
C_{\vartheta}(\Omega) \simeq C^1_{\vartheta}(\Omega) =  \left|\frac{F_\vartheta(\Omega)}{F_\vartheta(0)}\right|^2,
\label{eqn:FCF_gamma}
\end{equation}
where $F_\vartheta(\Omega)$ denotes the correlation function between field amplitudes, see Eq.~(\ref{eq:I}). Here, we assumed $\Omega\ll \omega$ and thus 
$\langle^{\omega+\Omega}I_{out}(\vartheta)\rangle\simeq\langle ^{\omega}I_{out}(\vartheta)\rangle=F_\vartheta(0)$ in the denominator of  Eq.~(\ref{eqn:FCF}).
Second, we use, as in Sec.~\ref{sec:MF_IntensityProp}, the \textit{approximation of minimal dephasing} in order to express $F_\vartheta(\Omega)$ in terms of ladder and crossed diagrams, see Eqs.~(\ref{eq:I_tot},\ref{gammaL_Om_tot},\ref{gammaC_Om_tot}).

We start by testing the validity of the approximate formula, Eq.~(\ref{eqn:gamma_final}), derived in Sec.~\ref{sec:MF_IntensityProp}.
Fig.~\ref{fig:Approx} confirms the very good agreement between Eq.~(\ref{eqn:gamma_final}) and the exact sum of ladder and crossed propagators given by Eqs.~(\ref{eq:I_tot},\ref{gammaL_Om_tot},\ref{gammaC_Om_tot}) for three different backscattering angles $\vartheta$. Furthermore, Fig.~\ref{fig:Approx} reveals that the frequency correlations $C^1_{\vartheta}(\Omega)$ sensitively depend on the backscattering angle $\vartheta$.

\begin{figure}
\includegraphics[width=0.49\textwidth]{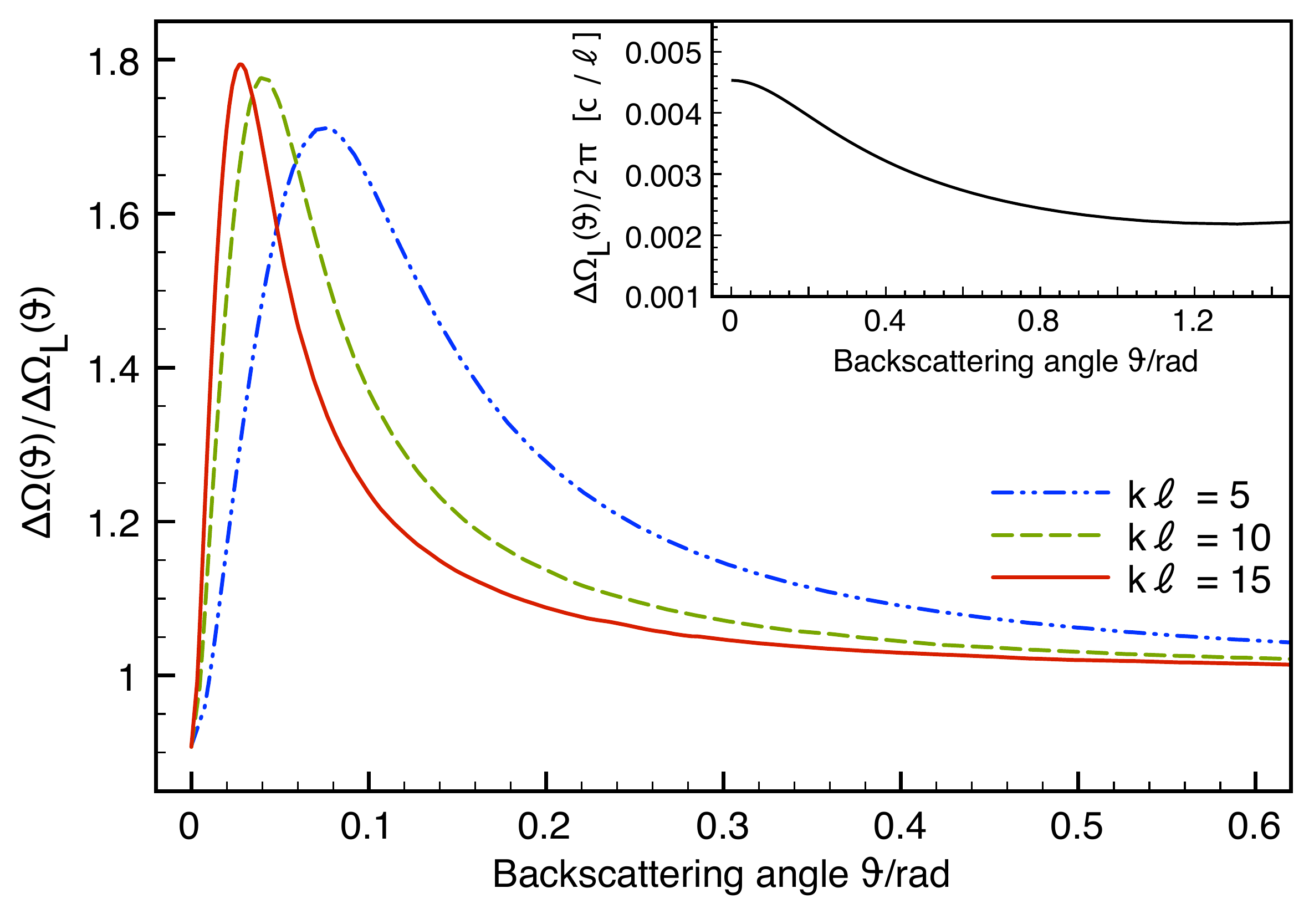}
\caption{Normalized width of the frequency correlation function $\Delta\Omega(\vartheta)/\Delta\Omega_L(\vartheta)$ as a function of the backscattering angle $\vartheta$ for different values for the disorder parameter $k\ell=15$ (red, solid line), $10$ (green, dashed line), and $5$ (blue, dot-dashed line) in weak disorder approximation. The data describes previous experimental results well in the regime of weak disorder $k\ell\gg 1$ but fails to describe the breakdown of the peak of the frequency broadening for stronger disorder (i.e., $k\ell\simeq 5$) reported in experiment \cite{muskens_angle_2011}. The inset shows the width $\Delta\Omega_L(\vartheta)$ in the case where crossed diagrams are excluded. Parameters: optical thickness $b=L/\ell=30$ and beam width  $\rho=100\ell$. \label{fig:DeltaOmega}}
\end{figure}

To further investigate this behaviour, we examine the width $\Delta\Omega(\vartheta)$ of the correlation function $C_{\vartheta}(\Omega)$, defined by $C_{\vartheta}(\Delta\Omega)
=C_{\vartheta}(0)/2$ as a function of the backscattering angle $\vartheta$ for three different values of the disorder parameter $k\ell$, see Fig.~\ref{fig:DeltaOmega}. In order to distinguish the influence of ladder and crossed diagrams, Fig.~\ref{fig:DeltaOmega} shows the normalized width $\Delta\Omega(\vartheta)/\Delta\Omega_L(\vartheta)$, where $\Delta\Omega_L(\vartheta)$ is the width of the frequency correlation function if only ladder diagrams are taken into account. Whereas the latter is independent of $k\ell$ and exhibits a slight monotonic decrease as a function of $\vartheta$ (see inset), a clear peak of the normalized width observed in  Fig.~\ref{fig:DeltaOmega} arises due to crossed propagation processes. According to Eq.~(\ref{eqn:gamma_final}), this behaviour can be interpreted in terms of the path length distributions $A^{(L,C)}_\vartheta(\mathcal{L})$: shorter paths lead to a broader frequency correlation function. In exact backscattering direction ($\vartheta=0$), the path lengths for ladder and crossed processes are identical -- except for single scattering which, as mentioned above, contributes only to ladder processes. For this reason, the normalized width starts slightly below the value 1 at $\vartheta=0$. At larger backscattering angles, longer path lengths contribute less to coherent backscattering (see also Appendix~\ref{sec:pathlength}), and the width of the frequency correlation function correspondingly increases. At the same time, however, the total  weight of crossed processes (integrated over all path lengths) decreases, such that the normalized width again approaches the value 1 for very large backscattering angles. In between, a maximum is found at $\vartheta\simeq \frac{1}{2k\ell}$, which approximately corresponds to the angular width of the coherent backscattering cone \cite{akkermans_coherent_1986}. The height of the peak slightly decreases for stronger disorder (i.e., lower values of $k\ell$). This can be traced back to the factor $h(\Omega)$ in Eq.~(\ref{eqn:gamma_final}) describing dephasing due to propagation outside the scattering medium (see Sec.~\ref{sec:MF_IntensityProp}), which becomes more relevant for larger backscattering angles.

For weak disorder (i.e., $k\ell\simeq 15$), the above results agree well with the experimental measurements \cite{muskens_angle_2011}.
For stronger disorder ($k\ell\simeq 5$), however, a drastic breakdown of the height of the peak was observed in the experiment \cite{muskens_angle_2011}, which is much more pronounced than the slight decrease predicted by Fig.~\ref{fig:DeltaOmega}.
We note that also the authors of \cite{muskens_angle_2011} are able to reproduce their measurements in the regime of weak disorder by theoretical calculations comparable to ours. In contrast to \cite{muskens_angle_2011}, however, our approach avoids additional approximations (such as, e.g., the diffusion approximation), and thus provides a more general and transparent theoretical interpretation. 
 
\section{Corrections to the frequency correlations for stronger disorder}
\label{sec:Corrections}

\begin{figure}
\includegraphics[width=0.35\textwidth]{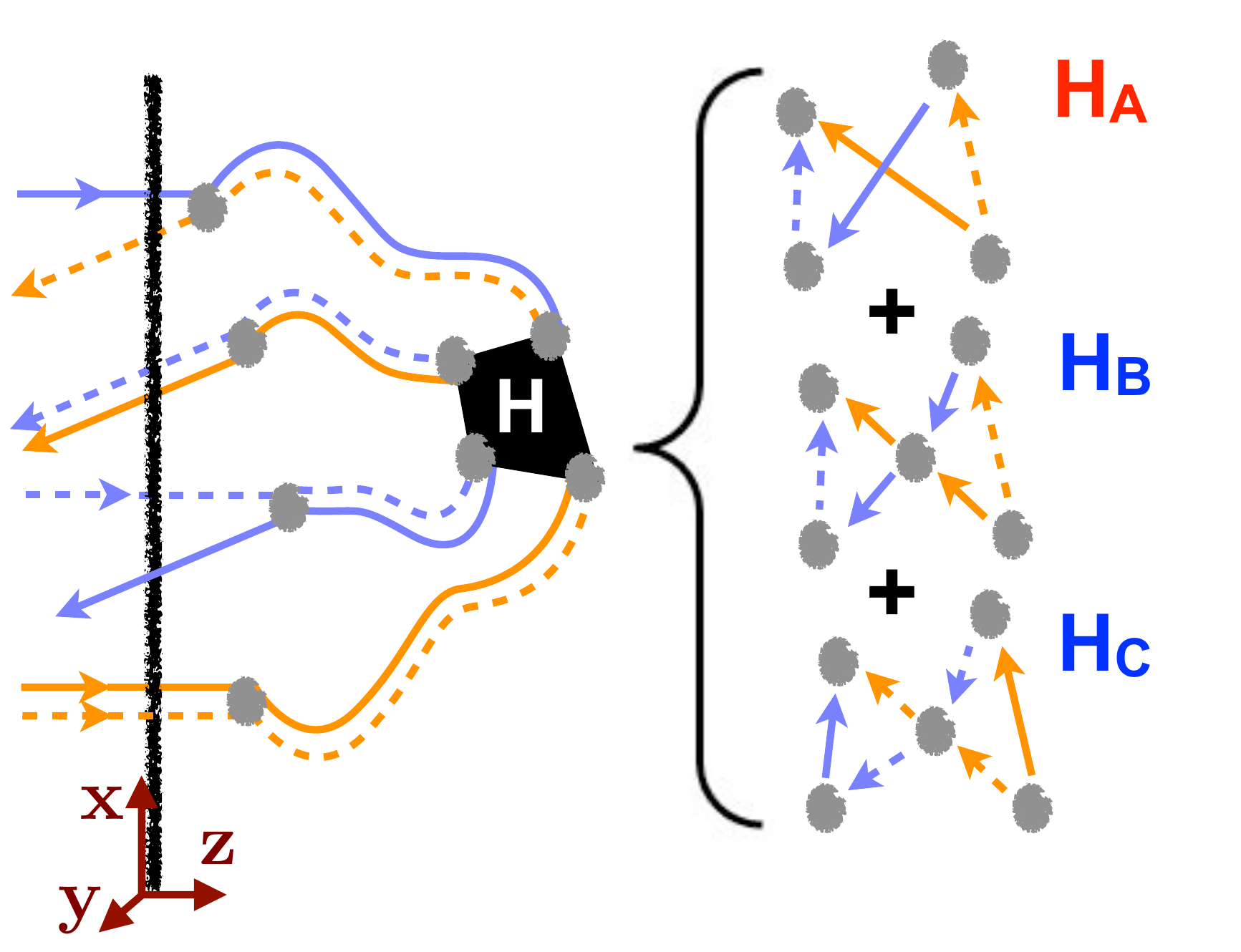}
\caption{Exemplary scattering scenario where the intensity propagators cross (black box $H$). 
The crossing  is described by a Hikami-box \cite{hikami_anderson_1981}  $H=H_A + H_B +H_C$. \label{fig:Hcrossing}}
\end{figure}

This discrepancy between the weak disorder description of frequency correlations
and the experimental observations in stronger disordered media calls for 
a study of possible extensions of the weak disorder theory. In this section, we first discuss corrections to the Gaussian approximation, which we find to be negligible for the setup under study. In a second approach, we extend the approximation of minimal dephasing by taking into account further scattering scenarios beyond the ladder and the crossed processes.

\subsection{Corrections to the Gaussian approximation}
\label{nonGauss}

The factorization of the four-field average in the course of the Gaussian approximation described in Sec.~\ref{sec:FCFweakdisorder} 
implies the underlying intensity propagators to follow independent scattering sequences that do not share any common scattering points. Such crossings between the intensity propagators, for which an exemplary trajectory is sketched in Fig.~\ref{fig:Hcrossing}, will induce non-Gaussian correlations among the fields.
We denote with $C_{\vartheta}^n(\Omega)$ a contribution to the frequency correlation function as defined in Eq.~(\ref{eqn:FCF}) with exactly $n-1$ points of crossing between the two mixed-frequency intensity propagators.
We first consider the case $n=2$ of a single crossing as follows:
In the same spirit as Berkovits describing angular correlations in reflection \cite{berkovits_long-range_1990}, we describe the crossing with a Hikami-box $H=H_A+H_B+H_C$  \cite{hikami_anderson_1981}  as depicted in Fig.~\ref{fig:Hcrossing} and the propagation of the intensity using diffusive propagators. In this frame, it is possible to evaluate the maximum of the $C^2$-contribution at zero frequency shift $(\Omega=0$) analytically \cite{knothe_frequency_2014}:
\begin{equation}
C_{\vartheta}^2(\Omega=0)=\frac{132}{49}\pi^3 \frac{\cos\vartheta}{(k\ell)^2 (S/\ell^2)},
\label{eqn:C2}
\end{equation}
where $S=\pi\rho^2$ denotes the surface of the Gaussian beam of width $\rho$. According to Eq.~(\ref{eqn:C2}), $C_{\vartheta}^2$ is a second order term in an expansion of the frequency correlation function in $1/(k\ell)$. The crucial dependence, however, stems from the behaviour  $C_{\vartheta}^2\propto \ell^2/S \propto (\ell/\rho)^{2}$. The appearance  of this factor can be explained as follows: if two photons enter the scattering medium at randomly chosen points within the transverse area $S\gg\ell^2$ of incidence, the probability of their crossing will be proportional to $1/S$. 
The same reasoning applies to cases with more crossings ($n>2$), which are of higher order in $1/(k\ell)$, but scale similarly with $1/S$: Once a single crossing has occurred, the probability of further crossings will be independent of $S$.
As already mentioned in Sec.~\ref{sec:MF_IntensityProp}, we assume $\rho\gg\ell$, since, in order to observe influences from crossed processes, for each scattering path a reversed counter path must exist. This is possible only if, both, the positions $\mathbf{r}_1$ and $\mathbf{r}_2$ of the first and of the last scattering event are covered by the laser beam, see also Eq.~(\ref{eqn:AApp}). The condition $\rho\gg\ell$, however, 
together with $S=\pi\rho^2$ 
proves contributions containing crossing points, i.e., all terms $C_{\vartheta}^n$ where $n>1$, to be negligible as compared to
$C^1_\vartheta(0)\equiv 1$, see Eq.~(\ref{eqn:FCF_gamma}), for the systems under study. Therefore, the Gaussian approximation $C_{\vartheta}\approx C_{\vartheta}^1$ remains a valid description of the frequency correlation function also in regime of stronger disorder.

\subsection{Influence of closed loops}
\label{subsec:closedloops}

\begin{figure}
\includegraphics[width=0.24\textwidth]{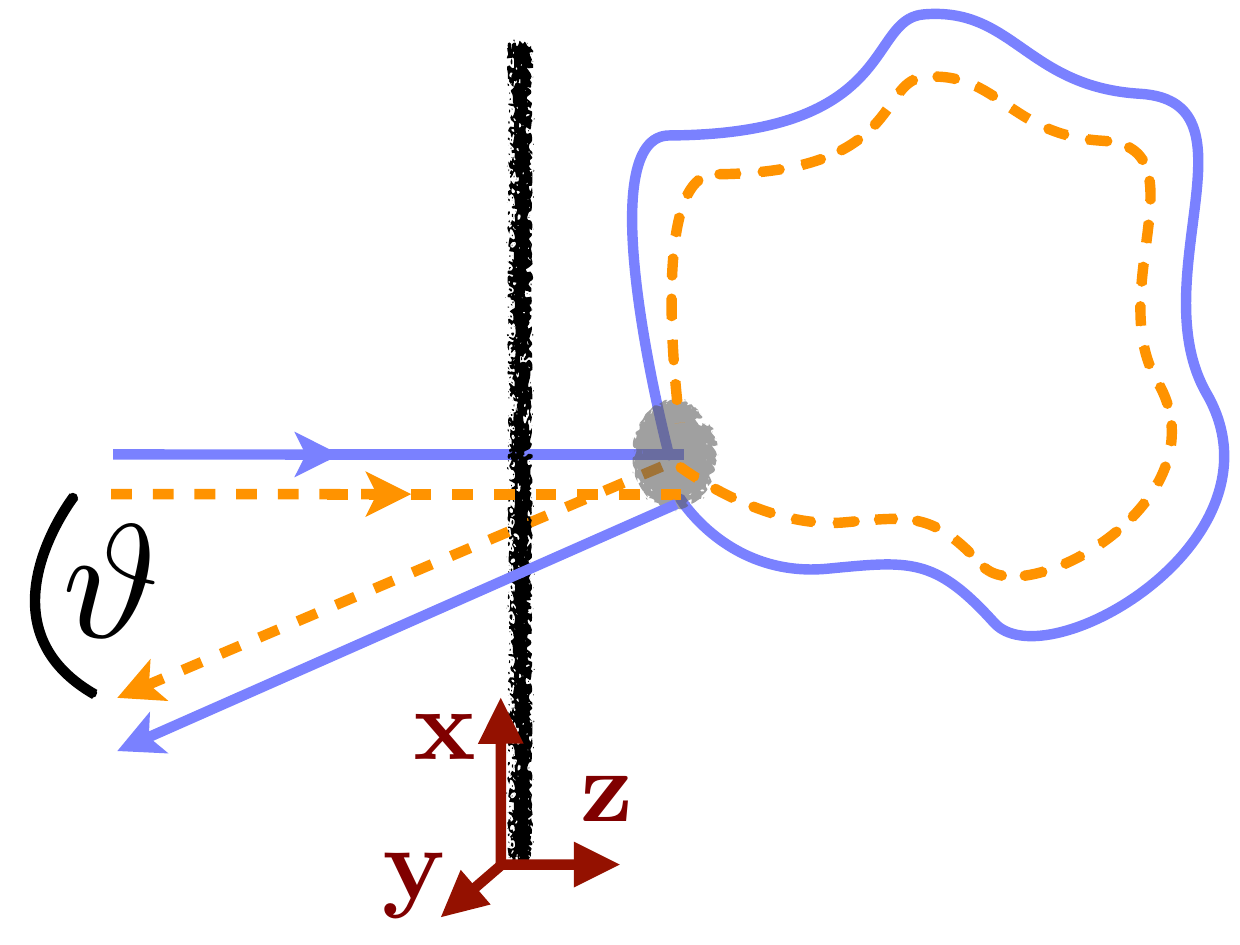}
\caption{Example of a recurrent scattering trajectory where the intensity propagates in a closed loop. We explore the consequences of these trajectories for the frequency correlation function as an extension of the weak disorder approximation of minimal dephasing, where only the contributions from ladder and crossed processes (see Fig.~\ref{fig:LundC}) are taken into account. \label{fig:RS_loop}}
\end{figure}

We now explore the consequences of additional scattering scenarios beyond the ladder and the crossed processes of Fig.~\ref{fig:LundC} for the frequency correlation function. The nature of the underlying scattering processes when the strength of the disorder approaches the threshold of Anderson localization has attracted attention recently in the context of experimental results \cite{aubry_recurrent_2014}: Changes in the behaviour of the angular correlations are reported as a consequence of so-called recurrent scattering trajectories, i.e., closed loop scattering scenarios where the initial and the final scatterer fall close to each other. Such a scattering process is sketched exemplarily in Fig.~\ref{fig:RS_loop}. The authors report a significant enhancement of the contributions from recurrent scattering trajectories to the backscattering signal as the value of the disorder parameter approaches the localization threshold. 

\begin{figure}
\includegraphics[width=0.49\textwidth]{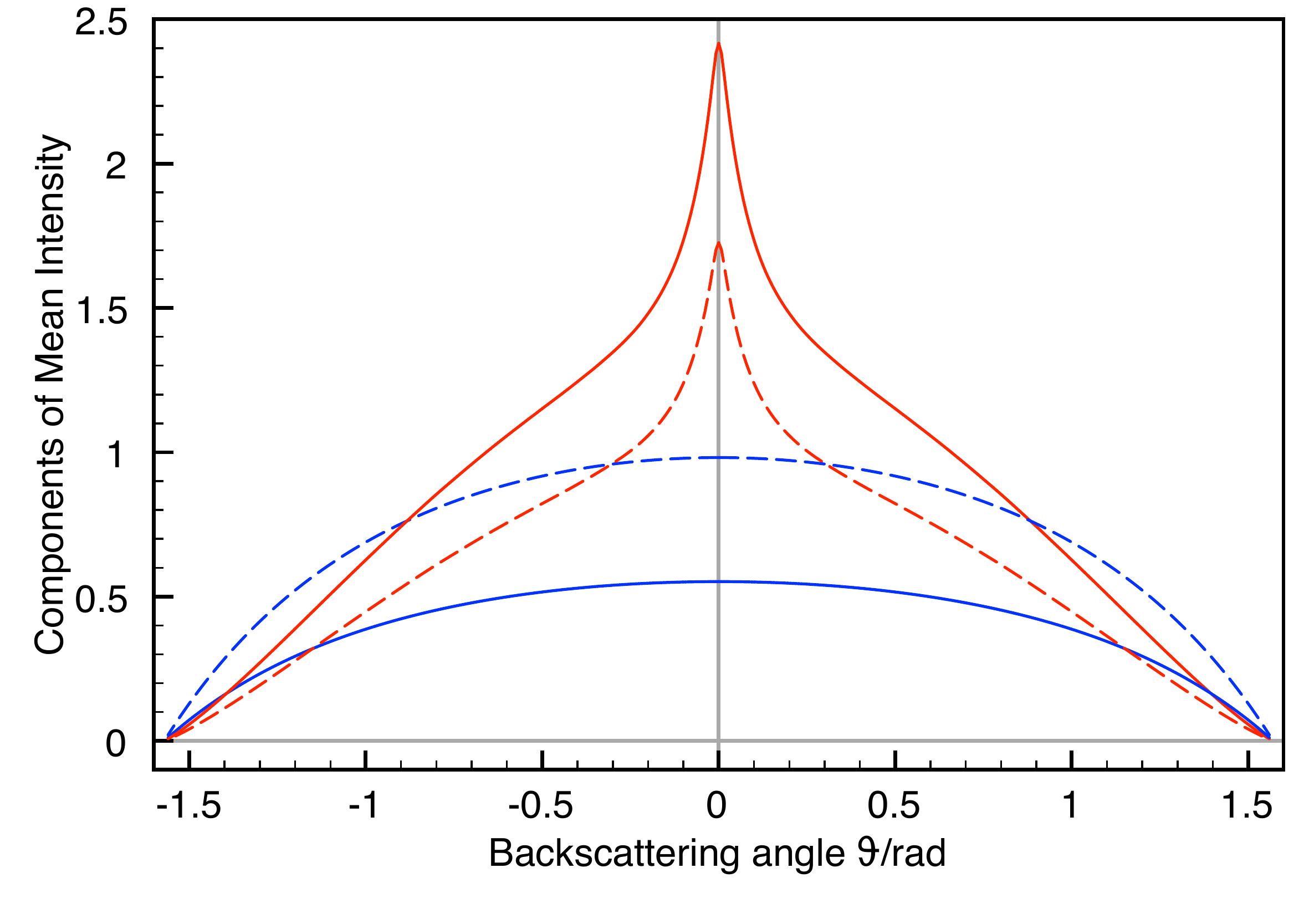}
\caption{Composition of the mean backscattered intensity $\gamma(\vartheta)=\gamma_{RS}(\vartheta)+\gamma_{CMS}(\vartheta)$ with different amounts of contributions $\gamma_{RS}(\vartheta)$ from Recurrent Scatterings (RS, blue) compared to contributions $\gamma_{CMS}(\vartheta)$ from Conventional Multiple Scatterings (CMS, red) exhibiting the characteristic coherent backscattering peak around $\vartheta\approx0$. We quantify the amount of RS contributions via the parameter $\Gamma=\gamma_{RS}(\vartheta=0)/\gamma_{CMS}(\vartheta=0)=0.23$ (solid lines) and $0.57$ (dashed lines). Contributions from single scattering events have been excluded. Remaining parameters as in Fig.~\ref{fig:Approx}.\label{fig:RS_Gamma}}
\end{figure}

In our Monte-Carlo simulations (see Appendix~\ref{sec:montecarlo}), we include a certain fraction of loop propagation processes into the scattering sequences in order to investigate their impact on the backscattering signal and the properties of the frequency correlation function. We account for the length of the scattering loop $\Lambda$ to be distributed as $1/\Lambda^2$, as it was observed in experiment \cite{aubry_recurrent_2014}, and the closed loops to exhibit angular properties like single scattering events.
The amount of contributions from closed loops as compared to the contributions from conventional multiple scattering events (ladder and crossed processes) in the backscattering signal is quantified by the parameter $\Gamma=\gamma_{RS}(\vartheta=0)/\gamma_{CMS}(\vartheta=0)$ as the ratio of mean backscattered intensity stemming from recurrent scattering events (RS) and from conventional multiple scattering processes (CMS), respectively, in exact backscattering direction $\vartheta=0$. This principle is demonstrated in Fig.~\ref{fig:RS_Gamma}, where we show the split-up between $\gamma_{RS}(\vartheta)$ and  $\gamma_{CMS}(\vartheta)$ for different values of $\Gamma$. Here, the backscattered intensity $^\omega I_{out}(\theta)$ is expressed as dimensionless quantity (\lq bistatic coefficient\rq\   \cite{akkermans_mesoscopic_2011}) defined as $\gamma(\theta)=4\pi R^2/(\pi \rho^2 I_0)\;^\omega I_{out}(\theta)$ with incident intensity $I_0=|\psi_0|^2$.

\begin{figure}
\includegraphics[width=0.49\textwidth]{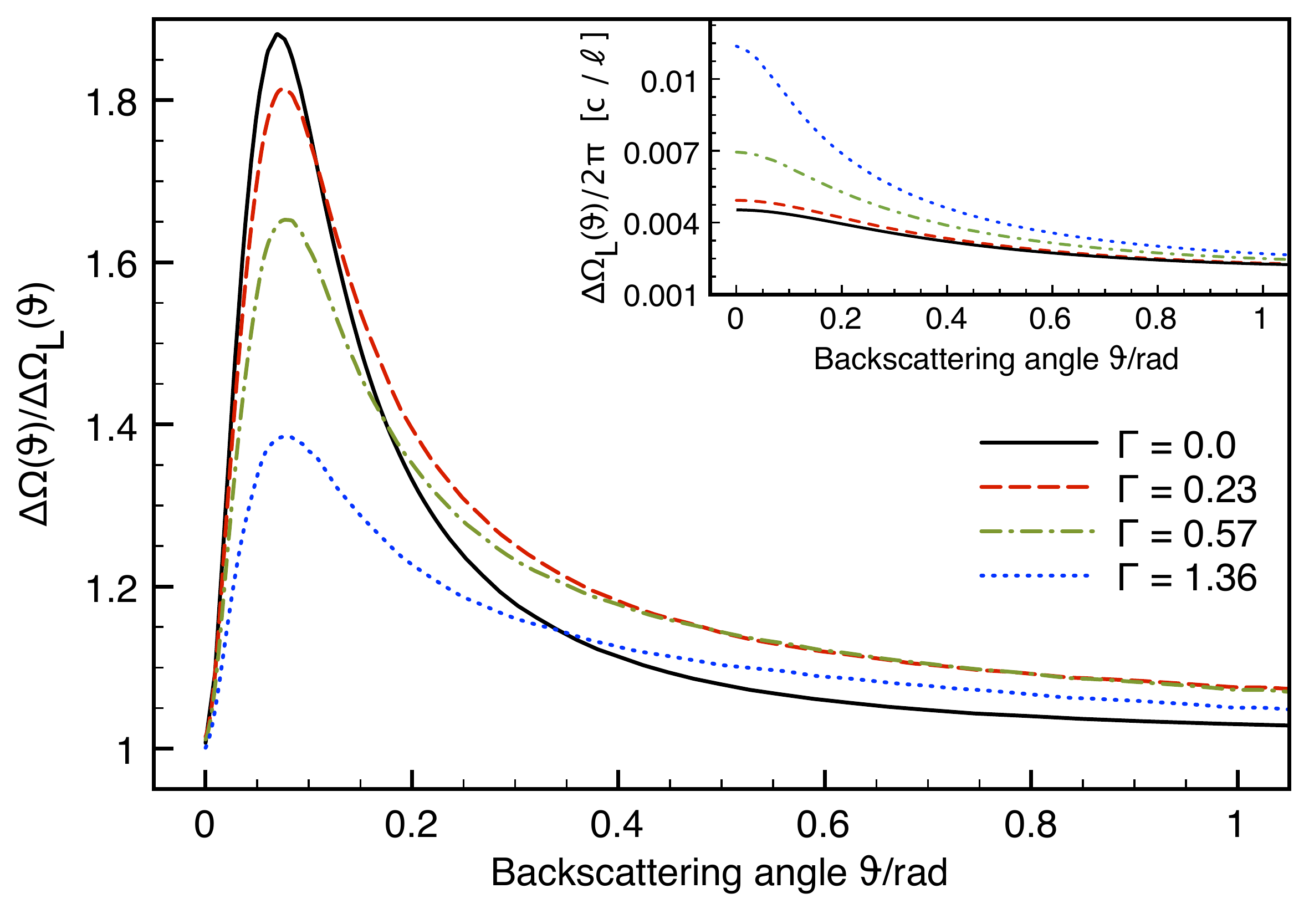}
\caption{Normalized frequency broadening  $\Delta\Omega(\vartheta)/\Delta\Omega_L(\vartheta)$ for different values of the parameter  $\Gamma=0$ (black, solid line), $0.23$ (red, dashed line), $0.57$ (green, dashed-dotted line), and $1.36$ (blue, dotted line) defining the amount of recurrent scattering trajectories to the backscattering signal, see Fig.~\ref{fig:RS_Gamma}. The peak in the normalized frequency broadening significantly decreases as the amount of recurrent scattering increases. 
The inset shows the width $\Delta\Omega_L(\vartheta)$ in the case where crossed diagrams are excluded. Remaining parameters as in Fig.~\ref{fig:Approx}.
\label{fig:RS_FCF}}
\end{figure}

After thereby having quantified the amount of recurrent scattering, its impact on the frequency correlation function is shown in Fig.~\ref{fig:RS_FCF}, where we plot the normalized width of the frequency correlation function $\Delta\Omega(\vartheta)/\Delta\Omega(0)$ for different values of $\Gamma$. We observe the peak height in the frequency broadening to decrease as the ratio of RS contributions in the backscattering signal increases. Since, as already discussed in Sec.~\ref{sec:FCFweakdisorder}, the peak can be traced back to the influence of crossed diagrams, this behaviour is consistent with the decreasing height of coherent backscattering cones observed in \ref{fig:RS_Gamma}.
Note that in the simulations behind Figs.~\ref{fig:RS_Gamma} and  \ref{fig:RS_FCF}, contributions from single scattering events are excluded from the backscattering signal to allow better comparability with experimental results \cite{muskens_angle_2011} (see also Appendix~\ref{sec:montecarlo}). 

These results give first evidence that the inclusion of scattering trajectories beyond the ladder and crossed diagrams indeed leads to a significant breakdown of the peak in the width of the frequency correlation function as a function of the backscattering angle $\vartheta$, in a similar way as it was observed experimentally \cite{muskens_angle_2011}. In our approach, we account for the occurrence of recurrent scattering trajectories \cite{aubry_recurrent_2014} on a phenomenological level, introducing them \lq by hand\rq\ into our Monte-Carlo simulations. In future work, it will be interesting to verify or check these results in the light of a more rigorous, consistent microscopic treatment of multiple scattering in the regime of stronger disorder.

\section{Conclusion and Outlook}

We theoretically investigated frequency correlations of light reflected from a random scattering medium. We established a direct relation between the frequency correlation function and the distribution of path lengths of multiple scattering sequences, which renders the path length distribution of the scattering sequences contributing in reflection directly accessible to experiment. Further, we extended these considerations to a description of intensity-intensity correlations in the regime of weak disorder where our model agrees with previous, experimental results \cite{muskens_angle_2011}. For the regime of stronger disorder, in which the same experiments report an unexpected and unexplained behaviour of the frequency correlation function, we investigated several extensions to the weak disorder model, in order to explain the discrepancy between experiments and theory in this regime. In particular, our results indicate that closed, recurrent scattering trajectories alter the properties of the backscattering signal in a way which points towards the experimentally observed behaviour.

Therefore, future studies of the properties of the frequency correlation function in disordered media should focus on confirming the impact of recurrent scattering events and describing their influence in more detail. On the side of experimental investigations, it would be of great interest to combine the experiments presented in \cite{muskens_angle_2011} and \cite{aubry_recurrent_2014}, in order to obtain joint information about the width of the frequency correlation function and the amount of closed loop trajectories contributing at a certain value of the disorder strength. From a theoretical point of view, the starting point could be two-fold: On the one hand, we lack a rigorous theoretical derivation predicting the fraction $\Gamma$ of closed loop trajectories as a function of the disorder parameter $k\ell$. This might be achieved using self-consistent approaches to localization \cite{vollhardt_diagrammatic_1980,skipetrov_dynamics_2006}. 
On the other hand, the diagrammatic representation of the recurrent scattering trajectories themselves is not yet fully clarified. 
In particular,  taking into account scattering sequences that return to a nearby scatterer (instead of exactly to the same scatterer, as we have assumed in the present paper) might lead to a different  angular behaviour of the underlying path length distribution. 

Starting from this point, the field is open for future investigations of the scattering dynamics close to the threshold of Anderson localization. The change of correlation properties in the angular \cite{aubry_recurrent_2014} or frequency \cite{hildebrand_observation_2014} domain have already proven to be a fruitful approach to obtain a better understanding of the underlying physical properties. Among these, frequency correlations appear especially promising due to their direct connection to dynamical properties and time scales. 
Therefore, future work on frequency correlations in random media will hopefully help to gain deeper and more profound insights about light transport and localization phenomena.

\begin{acknowledgments}
It is a pleasure to thank Otto L. Muskens for enlightening discussions. 
\end{acknowledgments}

\newcounter{zaehler}
\setcounter{zaehler}{\theequation}

\begin{appendix}
\section{Path length distributions}
\label{sec:pathlength}
\renewcommand{\theequation}{\arabic{equation}}
\setcounter{equation}{\thezaehler}

Here, we provide the exact definitions of the path and dephasing length distributions used in Eqs.~(\ref{eqn:Fourier}) and (\ref{eqn:gamma_final}).
They characterize the propagation of average intensity  for zero frequency shift  (i.e., $\Omega=0$). We start by expanding the propagator $P_L$ for $\Omega=0$ into a multiple scattering series formally obtained by iteration of Eq.~(\ref{eqn:Prop}):
\begin{eqnarray}
P_L(\mathbf{r}_i,\mathbf{r}_f) & = & u^2\delta(\mathbf{r}_i-\mathbf{r}_f)+u^4 P_0({\mathbf r}_i,\mathbf{r}_{f})+\sum_{n=3}^\infty u^{2n}\nonumber\\
& & \times\int d\mathbf{r}_{2}\dots d\mathbf{r}_{n-1}\prod_{j=1}^{n-1}P_0({\mathbf r}_j,\mathbf{r}_{j+1})\label{eq:A1}
\end{eqnarray}
where $\mathbf{r}_1\equiv \mathbf{r}_i$ and $\mathbf{r}_n\equiv \mathbf{r}_f$ in the second line. This allows us to write the propagator $P_L$ as an integral over an (unnormalized) path length distribution:
\begin{equation}
P_L(\mathbf{r}_i,\mathbf{r}_f) = \int d\mathcal{L} P^\mathcal{L}_L(\mathbf{r}_i,\mathbf{r}_f;\mathcal{L})
\end{equation}
where 
\begin{eqnarray}
P^\mathcal{L}_L(\mathbf{r}_i,\mathbf{r}_f;\mathcal{L}) & = & u^2\delta(\mathbf{r}_i-\mathbf{r}_f)\delta(\mathcal{L})\nonumber\\
& & +u^4 P_0({\mathbf r}_i,\mathbf{r}_{f})
\delta\left(\mathcal{L}-|\mathbf{r}_i-\mathbf{r}_f|\right)
\nonumber\\
&  &+\sum_{n=3}^\infty u^{2n}\int d\mathbf{r}_{2}\dots d\mathbf{r}_{n-1} \prod_{j=1}^{n-1}P_0({\mathbf r}_j,\mathbf{r}_{j+1})\nonumber\\
& & \times\delta\left(\mathcal{L}-\sum_{k=1}^{n-1} |\mathbf{r}_k-\mathbf{r}_{k+1}|\right).
\label{eqn:PLdistribution}
\end{eqnarray}
Inserting this into Eq.~(\ref{gammaL_Om_tot}), the joint distribution of path and enter/exit dephasing lengths defined by Eq.~(\ref{eqn:Fourier}) for the case of ladder propagation is obtained as follows:
\begin{eqnarray}
\mathfrak{A}_{L}(\mathcal{L},\lambda)  & = &  \frac{1}{(4 \pi R)^2} \iint d \mathbf{r}_{1}d \mathbf{r}_{2}\; e^{-\frac{(\mathbf{r}^{\bot}_1)^2}{\rho^2}}e^{-\frac{1}{\ell}(z_1+\frac{z_2}{\cos\vartheta})}\nonumber\\
& & \times P^\mathcal{L}_L(\mathbf{r}_{1},\mathbf{r}_{2};\mathcal{L})\delta\bigl(\lambda-\lambda_L(\mathbf{r}_{1}, \mathbf{r}_{2}, \vartheta)\bigr).\label{eq:plr1r2}
\end{eqnarray}
The corresponding expression $\mathfrak{A}_{C}(\mathcal{L},\lambda)$ for crossed processes is obtained in the same way from Eq.~(\ref{gammaC_Om_tot}). Attention must be paid, however, to the fact that single scattering processes are missing in $P^\mathcal{L}_C(\mathbf{r}_i,\mathbf{r}_f;\mathcal{L})$ as compared to the ladder case, i.e.,
$P^\mathcal{L}_C(\mathbf{r}_i,\mathbf{r}_f;\mathcal{L})=P^\mathcal{L}_L(\mathbf{r}_i,\mathbf{r}_f;\mathcal{L})-u^2\delta(\mathbf{r}_i-\mathbf{r}_f)\delta(\mathcal{L})$.
Integration over the dephasing length $\lambda$ yields the reduced path length distributions:
\begin{eqnarray}
A_{L}(\mathcal{L})  & = &  \frac{1}{(4 \pi R)^2} \iint d \mathbf{r}_{1}d \mathbf{r}_{2}\; e^{-\frac{(\mathbf{r}^{\bot}_1)^2}{\rho^2}}e^{-\frac{1}{\ell}(z_1+\frac{z_2}{\cos\vartheta})}\nonumber\\
& & \times P^\mathcal{L}_L(\mathbf{r}_{1},\mathbf{r}_{2};\mathcal{L}),\\
A_{C}(\mathcal{L}) &  = &  \frac{1}{(4 \pi R)^2} \iint d \mathbf{r}_{1}d \mathbf{r}_{2}\; e^{-\frac{(\mathbf{r}^{\bot}_1)^2+(\mathbf{r}^{\bot}_2)^2}{2\rho^2}}e^{-\frac{z_1+z_2}{2\ell}\left(1+\frac{1}{\cos\vartheta}\right)}\nonumber\\
& & \times e^{i \; ^{\omega}k\big((x_1-x_2)\sin\vartheta+(z_1-z_2)(1-\cos\vartheta)\big) }\nonumber\\
& & \times P^\mathcal{L}_C(\mathbf{r}_{1},\mathbf{r}_{2};\mathcal{L}).\label{eq:crossedpathlength}
\label{eqn:AApp}
\end{eqnarray}
From the above expressions, 
we first observe that, for exact backscattering $\vartheta=0$ and large beam widths $\rho\to\infty$, the expressions for $A_{L}(\mathcal{L})$ and $A_{C}(\mathcal{L})$ are identical (except for the single scattering contribution). For non-zero backscattering angles $\vartheta>0$, a non-vanishing complex phase factor is present in the crossed amplitude $A_{C}(\mathcal{L})$. Upon disorder average, this leads to the suppression of long light paths for the crossed processes with rising angle mentioned in Sec.~\ref{sec:MF_IntensityProp}. In a similar way, crossed processes are also suppressed if the  width $\rho$ of the laser beam becomes smaller than the transverse distance $|\mathbf{r}^{\bot}_1-\mathbf{r}^{\bot}_2|$ between the first and the last scattering event in Eq.~(\ref{eq:crossedpathlength}), which is typically comparable to the mean free path $\ell$.

\section{Monte-Carlo simulations}
\label{sec:montecarlo}

In this appendix, we sketch the Monte-Carlo algorithm used for our simulations of the 
backscattered radiation. 
In the regime of weak disorder, where we apply the approximation of minimal dephasing described in Sec.~\ref{sec:MF_IntensityProp}, we insert the representation of the propagator as given in Eq.~(\ref{eq:A1}) into  Eqs.~(\ref{gammaL_Om_tot},\ref{gammaC_Om_tot}) to obtain multi-dimensional integrals $\int d\mathbf{r}_1\dots d\mathbf{r}_n$ over the positions of scattering events. These can be solved numerically by virtue of an iterative procedure in which each single propagation step is simulated by drawing random numbers from appropriate distributions determining first the initial penetration depth and subsequently the angular orientation and the length of each respective step \cite{labeyrie_coherent_2003}. Averaging over different realizations of the disorder is then achieved by repeating the algorithm numerous times. 
The number of iterations $N$  for the simulations amounted to $N=10^7$ for the data presented in Figs.~\ref{fig:Approx} and \ref{fig:DeltaOmega}, $N=10^6$ in Fig.~\ref{fig:RS_Gamma} and $N=10^5$ in Fig.~
\ref{fig:RS_FCF}. At these numbers, the statistical error on the computed quantities was assured to be smaller than visible on the scale shown.

This scheme is extended to the regime of stronger disorder by further taking into account recurrent scattering (RS) trajectories. Implementing RS paths into the scattering sequences in a consistent and flux conserving way is achieved as follows:  First, we define a weighting factor $p$ as the probability that propagation occurs along a closed loop. Choosing different values for $p$ corresponds to varying $\Gamma$ in Figs.~\ref{fig:RS_Gamma} and \ref{fig:RS_FCF}. In accordance with experiment \cite{aubry_recurrent_2014}, the length of the loop is distributed as $P_{RS}(\Lambda)=2\ell/\Lambda^2$ for $\Lambda\geq 2\ell$ (and $0$ otherwise, i.e., the minimal length of a loop is assumed as $2\ell$). Furthermore, the two fields (solid and dashed lines in Fig.~\ref{fig:RS_loop}) can complete the loop either in the same sense (clockwise or counter-clockwise) or in the opposite sense, giving rise to equally contributing ladder and crossed diagrams. In total, the recurrent scattering contribution thus results by modifying ladder and crossed propagators as follows: $P^\mathcal{L}_L(\mathbf{r}_i,\mathbf{r}_f;\mathcal{L})=P^\mathcal{L}_C(\mathbf{r}_i,\mathbf{r}_f;\mathcal{L})=\frac{p}{2}u^2\delta(\mathbf{r}_i-\mathbf{r}_f)P_{RS}(\Lambda)$ and inserting these expressions into Eqs.~(\ref{eq:plr1r2}) and (\ref{eqn:Fourier}). At the same time, the conventional multiple scattering contributions are multiplied by the factor $1-p$. In order to ensure flux conservation \cite{knothe_flux_2013}, the conventional ladder component is multiplied by an additional factor which we determine numerically by integrating the flux over all angles.  

For comparison with experiments on light scattering, we furthermore account for the polarization degree of freedom as follows: single scattering is filtered out  in the helicity conserving and linear non-conserving detection channel, whereas scattering processes of higher order are assumed to be equally split into the respective conserving or non-conserving channel (for linearly or circularly polarized light). Likewise, crossed propagation processes fully contribute to the helicity conserving or linear conserving channels, whereas they are filtered out in the respective orthogonal channels. Naturally, the condition of flux conservation must hold for the sum of two orthogonal channels. Under these premises, the results of Fig.~\ref{fig:RS_Gamma} refer to the helicity conserving channel (where crossed processes contribute and single scattering is filtered out). Likewise, the data for $\Delta\Omega(\theta)$ in Fig.~\ref{fig:RS_FCF} also refers to the helicity conserving channel, whereas $\Delta\Omega_L(\theta)$ (inset) is evaluated in the linear non-conserving channel (only ladder processes, single scattering filtered out).

\end{appendix}

\end{document}